\documentclass[aps]{revtex4}

\usepackage{graphicx}
\usepackage{dcolumn}
\usepackage{bm}
\usepackage{amsmath}

\begin{document}
\draft
\newcommand{\eqbeg}{\begin{equation}}
\newcommand{\eqend}{\end{equation}}
\newcommand{\mone}{\mbox{m}_1}
\newcommand{\mtwo}{\mbox{m}_2}
\newcommand{\mthree}{\mbox{m}_3}
\newcommand{\ddt}{\frac{d}{dt}}
\newcommand{\toh} {t_0}
\newcommand{\lone} {L$_1$ }
\newcommand{\cone} {C$_1$ }
\newcommand{\czero} {C$_0$ }
\newcommand{\rone} {R$_1$}
\newcommand{\zone} {Z$_1$}
\newcommand{\ztwo} {Z$_2$}
\newcommand{\cnb} {C$\nu$B }

\setlength{\tabcolsep}{.1in}

\title{  Comment on ``Fundamental Impossibility of a Superradiant Neutrino Laser'', by Y-K. Lu, H. Lin, and W. Ketterle , Phys. Rev. Lett. 137, 101804  (2026) and ``Can Bose-Einstein Condensates Enhance Radioactive Decay?'', by H. Lin, Y-K. Lu, and W. Ketterle, Phys. Rev. Lett. 137, 101805 (2026) }
\author{Ralph DeVoe}
\address{Department of Physics, Stanford University, Stanford, CA 94305}
\date{\today}
   
\maketitle


Recently three Letters \cite{ketterle, ketterle_02, formaggio} have been published on  the possibility of making a superradiant neutrino laser in a Bose-Einstein condensate (BEC). Formaggio et. al. \cite{formaggio} proposed  that a condensate of $^{83}$Rb could emit neutrinos with a rate $> 1000$ times the single atom decay rate due to superradiance. Ketterle et. al. \cite{ketterle} showed that the process was forbidden by fermi statistics due to destructive interference of transition amplitudes or a Pauli blockade. A further study \cite{ketterle_02} considered neutrino superradiance as part of a general study of BEC's and concluded that it was also forbidden for other reasons. There is, however,  a simple but fundamental explanation of why neutrino superradiance cannot occur in the lab which is not mentioned in any of these papers: the weak interaction is so weak,  with  cross-sections in the $10^{-40}$ to $10^{-44}$ cm$^2$ range, that the correlations necessary for superradiance cannot occur.  This depends on an understanding of how superradiance is generated rather than on the details of any experiment.

The relevance of correlations to Dicke superradiance \cite{dicke} can be shown by using the approach of Gross and Haroche \cite{haroche}. Defining $\Gamma$ as the single atom spontaneous emission rate the  superradiant decay rate is
\eqbeg
                \Gamma < J_{+} J_{-}> = \Gamma  <(  \sum_{i=1}^{i=N} {J_+^i }  ) (  \sum_{k=1}^{k=N} J_-^{k} )> =
                     \Gamma  <\sum_{i} J_+^i J_{-}^i> + \Gamma < \sum_{i} \sum _{k \ne i } J_+^i J_-^k > 
\eqend 
where $J_+$ and $J_-$ are the collective operators and and $J_+^i$, and  $J_-^i$ are those for the i-th atom. The single sum on the right represents independent spontaneous emission while the double sum gives the $N(N-1)$ factor in the superradiant decay rate.  As noted in \cite{haroche} the terms that are off-diagonal in i and k  dominate the decay rate. That is, the enhanced decay rate of superradiance is entirely due to \it correlations \rm between the raising operator of i-th atom and the lowering operator of the k-th atom. Since the atoms are initially uncorrelated,  the i-th and the k-th atoms must exchange information during the decay. This means that spin-spin interactions must occur at a rate large compared to the single atom rate $\Gamma$. This is established through the coherence requirement  kR $<$1 which places the atoms less than $  \lambda / 2 \pi$ apart. Since the cross-section for photon absorption \cite{mandel} is on the order of $\lambda^2$, the coherence requirement is also a requirement that the sample be optically dense. In the limiting case in which the Dicke states $|J,m_J>$ represent the system, a spontaneous photon will be coherently scattered many times before leaving the sample.  This is how the correlations arise.  The importance of the cross section in superradiance is particularly clear in small sample superradiance , for example, in two trapped ions \cite{dev_sr}.  There is no laboratory system which is optically dense for neutrinos since their mean free path is at least 10$^{20}$ cm.  A neutrino emitted in the lab has essentially zero chance of interacting with another part of the experiment and superradiance cannot occur, regardless of its size or shape and whether it is a BEC or normal matter.

\end{document}